\documentclass[aps,twocolumn,showpacs,preprintnumbers,amsmath,amssymb,prl,10pt,floatfix,superscriptaddress]{revtex4}
\usepackage{graphicx,amsbsy}

\begin{document}

\title{
Correlation tuned cross-over between thermal and nonthermal states following ultrafast transient pumping
}

\date{\today}

\author{B. Moritz}
\email{moritz.physics@gmail.com}
\affiliation{Stanford Institute for Materials and Energy Sciences, SLAC National Accelerator Laboratory, Menlo Park, CA 94025, USA}
\affiliation{Department of Physics and Astrophysics, University of North Dakota, Grand Forks, ND 58202, USA}
\affiliation{Department of Physics, Northern Illinois University, DeKalb, IL 60115, USA}

\author{A. F. Kemper}
\author{M. Sentef}
\author{T. P. Devereaux}
\affiliation{Stanford Institute for Materials and Energy Sciences, SLAC National Accelerator Laboratory, Menlo Park, CA 94025, USA}

\author{J. K. Freericks}
\affiliation{Department of Physics, Georgetown University, Washington, DC 20057, USA}

\begin{abstract}
We examine electron-electron mediated relaxation following excitation of a correlated system by an ultrafast electric field pump pulse.  The results reveal a dichotomy in the temporal evolution as one tunes through a Mott metal-to-insulator transition:  in the metallic regime relaxation can be characterized by evolution toward a steady-state electronic distribution well described by Fermi-Dirac statistics with an increased effective temperature; however, in the insulating regime this quasithermal paradigm breaks down with relaxation toward a nonthermal state with a more complicated electronic distribution that does not vary monotonically as a function of energy.  We characterize the behavior by studying changes in the energy, photoemission response, and electronic distribution as functions of time.  Qualitatively these results should be observable on short enough time scales that the electrons behave like an isolated system not in contact with additional degrees of freedom which can act as a thermal bath.  
Importantly, proper modeling used to analyze experimental findings should account for this behavior, especially when using strong driving fields or studying materials whose physics may manifest the effects of strong correlations.
\end{abstract}
 

\maketitle

Optical reflectivity,\cite{Bovensiepen_JPCM_2007,Rini_Nature_2007,Fausti_Science_2011,Wall_NatPhys_2011,Dal_Conte_Science_2012} photoemission spectroscopy, \cite{Bovensiepen_JPCM_2007,Perfetti_PRL_2006,Cavalieri_Nature_2007,Perfetti_PRL_2007,Schmitt_Science_2008,Graf_NatPhys_2011,Ishida_ScientificRep_2011,Cortes_PRL_2011,Rettig_PRL_2012,Sobota_PRL_2012,Smallwood_Science_2012} and resonant x-ray scattering\cite{Forst_PRB_2011,Ehrke_PRL_2011,Lee_NatureComm_2012} are equilibrium methods which in the time domain are ideally suited to studying dynamics of novel ordered phases or collective excitations.\cite{Rini_Nature_2007,Fausti_Science_2011,Wall_NatPhys_2011,Dal_Conte_Science_2012,Perfetti_PRL_2006,Perfetti_PRL_2007,Schmitt_Science_2008,Graf_NatPhys_2011,Cortes_PRL_2011,Rettig_PRL_2012,Sobota_PRL_2012,Smallwood_Science_2012,Forst_PRB_2011,Ehrke_PRL_2011,Lee_NatureComm_2012}  On sufficiently short time scales, the initial recovery in these systems following an ultrafast pump pulse should be dominated by 
electron-electron scattering which on its own can drive the system into a new steady-state.  Conventional analysis has been based on a quasithermal paradigm (``hot electron'' or multi-temperature models);\cite{Anisimov_JETP_1974,Allen_PRL_1987}  however, there have been few tests of the validity of its underlying assumptions as a function of the strength of electronic correlations,\cite{Eckstein_PRB_2011} in particular as one tunes between the two regimes of a metal-to-insulator transition (MIT).\cite{Imada_RMP_1998}

The MIT driven by electronic correlations usually is accompanied by a number of interesting ordering phenomena among the spin, charge, and orbital degrees of freedom in a material.  An understanding of the key physics which leads to these emergent phases is often at the heart of pump-probe experiments in condensed matter systems, including high-$T_{c}$ cuprate superconductors,\cite{Lee_RMP_2006} nickelates, manganites, ruthenates, vanadates,\cite{Imada_RMP_1998} and even organic materials.\cite{Lefebvre_PRL_2000,Powell_RPP_2011,Ardavan_JPSJ_2012}  A number of experimental parameters can be used to tune across the MIT including doping and chemical substitution, pressure, and applied fields.  What can be learned about the underlying physics leading to these phases as a function of these key parameters requires an understanding of the proper paradigm in which to ask the relevant questions and conduct analysis of experimental data.  This is in addition to what can be learned by tuning the interaction parameters 
of model systems simulated in fermionic or bosonic cold atom mixtures and performing the experimental equivalent of time-resolved, pump-probe measurements.\cite{Jin_Nature_2008,Jordens_Nature_2008,Trotzky_NatPhys_2012}

In this Letter we discuss the evolution of an electronic system described in equilibrium by a simple Hamiltonian representing a correlated electronic system possessing a MIT tuned by the strength of the Coulomb repulsion $U$. In order to avoid approximate treatments of either correlations or applied fields, we chose to study the spinless Falicov-Kimball model\cite{Falicov_PRL_1969} whose effective Hamiltonian is given by
\[
H = -\frac{t^{*}}{2\sqrt{d}}\sum_{\left<ij\right>}(c_{i}^{\dag}c_{j} + h.c.)-\mu\sum_{i}c_{i}^{\dag}c_{i}+U\sum_{i}w_{i}c_{i}^{\dag}c_{i}.
\]
This model describes itinerant conduction electrons hopping between lattice sites with an energy $t^{*}$ and chemical potential $\mu$ that experience Coulomb repulsion with another species of localized electrons with an occupation $w_{i}$ on each site.  This model can be tuned through a Mott MIT at half-filling by adjusting the electron-electron interaction $U$ with $U_{c}=\sqrt{2}t^{*}$.  Here we wish to understand, in general terms, how electron dynamics are affected by both strong fields and strong correlations; therefore, our exact treatment for this model is of more general interest than using a materials-specific Hamiltonian that may require a number of approximations to affect a full solution in the time-domain.

\begin{figure}[!t]
\includegraphics[width=\columnwidth]{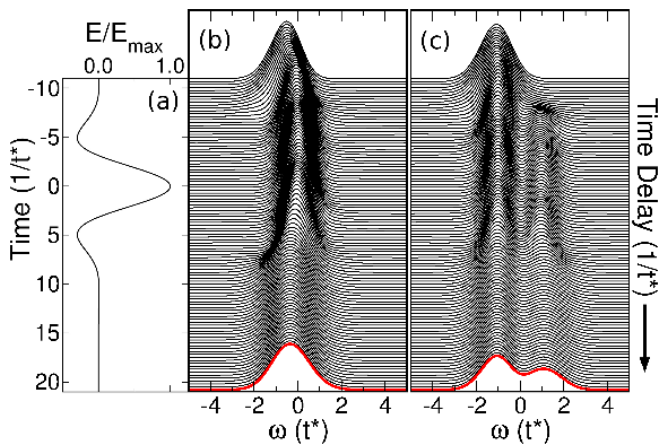}
\caption{(a) The electric field pump pulse that leads to the temporal evolution shown in panels (b) and (c) has a maximum intensity $E_\textrm{max}$=24$E_\textrm{o}$ (normalized in the panel), modulation frequency $\omega_\textrm{p}=0.5t^{*}$, and characteristic width $\sigma_\textrm{p}=5/t^{*}$.  (b) \& (c) Time-resolved pump-probe photoemission response for (b) metallic ($U=0.5t^{*}$) and (c) insulating ($U=2t^{*}$) systems determined for a probe pulse with characteristic width $\sigma_\textrm{b}=2/t^{*}$.\label{fig:1}}
\end{figure}

We model the transient pump pulse as a spatially uniform, harmonic, electric field with a Gaussian envelope [see the temporal profile shown in Fig.~\ref{fig:1}(a)] incorporated via the Peierls' substitution\cite{Peierls_ZPhys_1933} in the Hamiltonian gauge.  The temporal evolution is simulated using an exact, nonequilibrium formulation of DMFT.\cite{Georges_RMP_1996,Freericks_PRL_2006,Freericks_PRB_2008,Eckstein_PRL_2008,Eckstein_PRL_2009}  Following standard convention, the energy unit is taken to be $t^{*}$ throughout this work and the standard unit for time is $1/t^{*}$.  Conversion to physical units and the effective electric field scale, denoted $E_\textrm{o}$, as well as details about the evaluation of an effective quasithermal response, can be found in the Supplementary Material.  

Figures~\ref{fig:1}(b) and (c) show the characteristic photoemission response\cite{Eckstein_PRB_2008,Freericks_PRL_2009,Eckstein_PRB_2010} as a function of time delay for representative metallic and insulating systems photoexcited by the pump-pulse shown in Fig.~\ref{fig:1}(a).  The width of the photoemission probe pulse influences both the temporal resolution and energy resolution of the resulting spectrum, chosen here to strike a balance between the two.  

For metallic correlations [Fig~\ref{fig:1}(b), $U=0.5t^{*}$] the pump pulse narrows and shifts the response toward the equilibrium Fermi level ($\omega=0t^{*}$).  Following the pump pulse, we observe a rapid relaxation toward a significantly broader spectral distribution characteristic of a new steady-state.  For stronger correlations on the insulating side of the Mott MIT [Fig~\ref{fig:1}(c), $U=2t^{*}$] the pump pulse narrows the response below the equilibrium Fermi level and transfers spectral weight across the insulating Mott gap (centered at the Fermi level).  Similar to the behavior observed for weak correlations, we find a broader spectral distribution following the transient pump, although in this case a significant remnant spectral weight transfer across the gap persists, even in the long-time steady-state.  This rapid evolution following the decay of the pump pulse can be attributed to relaxation mediated by electron-electron scattering.

\begin{figure}[!t]
\includegraphics[width=\columnwidth]{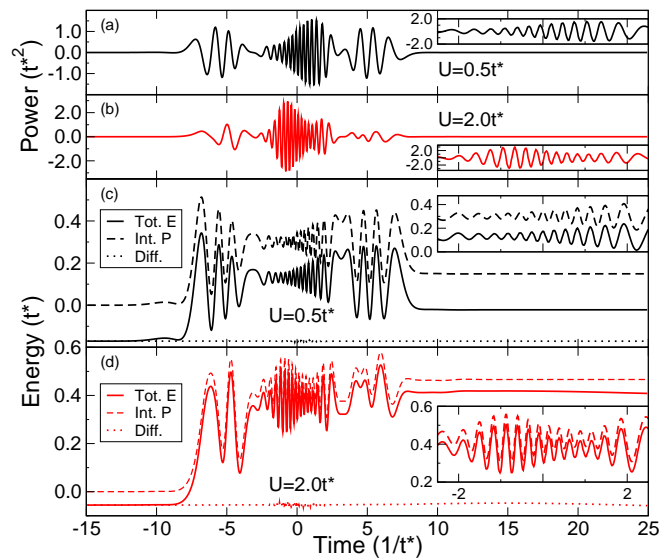}
\caption{Instantaneous power delivered to (a) metallic ($U=0.5t^{*}$) and (b) insulating ($U=2t^{*}$) systems discussed in Fig.~\ref{fig:1}.  (c) \& (d) As a closed system, the change in total energy may be determined from the local propagator (solid lines) or integrating the instantaneous power (dashed lines). The constant offset (dotted lines) represents the initial, equilibrium value in each case.  The insets highlight times near the center of the pump pulse between $-2.5/t^{*}$ and $2.5/t^{*}$ that show rapid variation in each quantity.\label{fig:2}}
\end{figure}

An increase in the effective temperature would naturally lead to broader features and a redistribution of weight across the insulating Mott gap, as the spectral function in equilibrium is temperature independent for this model. The highlighted (red) traces in Fig.~\ref{fig:1} can be used to assess whether the steady-state is representative of the system in equilibrium at an elevated, effective temperature.  However, as we now will show, our results highlight a distinct dichotomy in the temporal evolution of as one tunes across the MIT.  In the metallic regime the quasithermal picture remains valid with electron-electron mediated relaxation characterized by evolution toward an effectively thermal steady-state.  However, tuning correlations across the MIT causes a breakdown in the quasithermal paradigm in the insulating regime with clearly nonthermal relaxation even at long times.

We first determine the appropriate temperature based on the change in total energy of the system.\cite{Turkowski_PRB_2006}  Figure~\ref{fig:2} shows the power delivered to the system by the pump pulse as well as the time-dependent total energy for the systems discussed in Fig.~\ref{fig:1}.  Figures~\ref{fig:2}(c) and (d) show a comparison between the total energy and the change in total energy determined by integrating the instantaneous power.  As one can see, the simulation properly accounts for the energy delivered by the pump pulse with small deviations near time $0/t^{*}$ (near the center of the pulse) where the power (and instantaneous current) changes rapidly.  The offset simply reflects the initial energy in equilibrium.  Converting the total energy at long times to an effective temperature (assuming validity of the ``hot electron'' model), we find an increase of $\sim8$ times for the weakly correlated, metallic system and $\sim43$ times for the strongly correlated, insulating system for identical 
driving fields.
 
\begin{figure}[!t]
\includegraphics[width=\columnwidth]{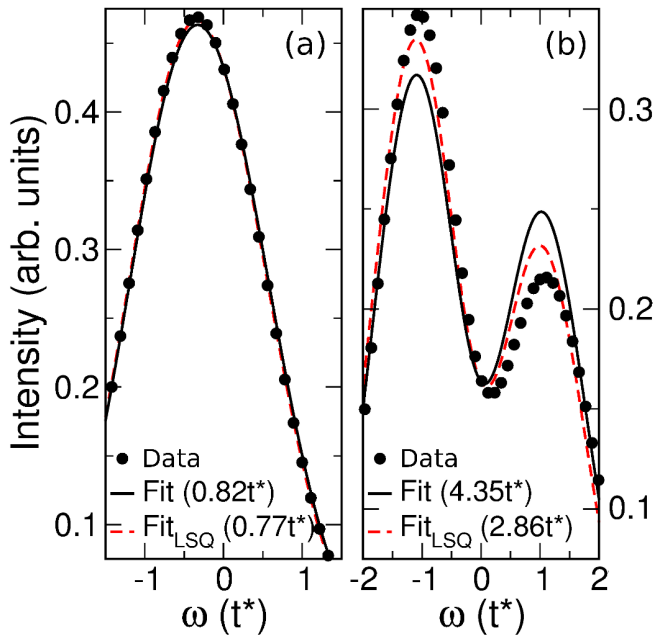}
\caption{Quasithermal fits of the steady-state photoemission response (characteristic probe width $\sigma_\textrm{b}=2/t^{*}$) for (a) metallic ($U=0.5t^{*}$) and (b) insulating ($U=2t^{*}$) systems, respectively, with the effective temperature shown in parenthesis.  The fits (solid black) are determined by extracting the temperature from the total energy, while the best-fit quasithermal response (dashed red) has been determined by the least-squares method (LSQ).\label{fig:3}}
\end{figure}
 
Figure~\ref{fig:3} shows the result of quasithermal fits to the photoemission response in both [(a), $U=0.5t^{*}$] the metallic and [(b), $U=2.0t^{*}$] insulating regimes.  In each case the characteristic probe width is fixed as in Fig.~\ref{fig:1}.  In the weakly correlated metal the simulated response closely matches that derived from the quasithermal fits with only small differences which grow upon increasing the pump strength and frequency as would befit simple expectations (see the Supplementary Material).  In this case the ``best fit'' also has been determined from a simple least-squares fit (LSQ), but with a similar conclusion.  

In contrast, the steady-state response in the insulating regime shows significant deviation from a quasithermal best fit, determined using either method, clearly indicating a breakdown in the ``hot-electron'' model.  Issues associated with convergence of our self-consistent method restrict our studies to relatively strong driving fields; however, we can infer that similar observations should hold as the strength of the transient field is reduced, although the deviation from the quasithermal paradigm is likely to be noticeable only in the tail of the spectral function across the Fermi level, or, equivalently, the extracted, effective Fermi-Dirac distribution.  Observations of nonthermal behavior already have been made experimentally for comparatively weaker driving fields.\cite{Lee_NatureComm_2012}   

We also examine the instantaneous electronic distribution providing a snapshot of the response to the applied electric field and subsequent relaxation due to electron-electron scattering.  Consider a simple noninteracting band metal.  The initial distribution follows usual Fermi-Dirac statistics $f(\varepsilon_\mathbf{k}-\mu)$ with preferential occupation of the lower-energy states according to temperature.  When driven by an electric field, an electron initially at momentum $\mathbf{k}$ shifts to $\mathbf{k}-e\mathbf{A}(t)$ (the standard Peierls' substitution).  This behavior can be extracted directly from the gauge-invariant lesser Green's function for the system.  

\begin{figure}[!t]
\includegraphics[width=\columnwidth]{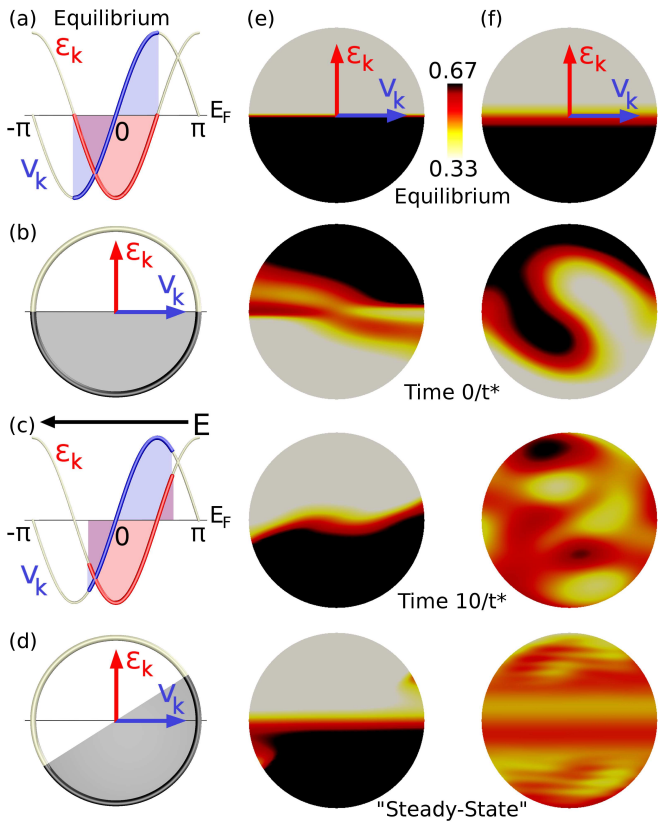}
\caption{(a)-(d) A simple cartoon depicting the influence of an applied electric field on a noninteracting one-dimensional band metal at half-filling.  (a) \& (c) The band energy ($\varepsilon_\mathbf{k}$, red highlight) and normalized band velocity (v$_\mathbf{k}$, blue highlight) distributions in equilibrium and under the influence of an applied electric field, respectively.  In one-dimension each band energy is associated with two band velocities (right (+) and left (-)).  In higher dimensions a distribution of velocities is associated with each band energy.  (b) \& (d) The equivalent distribution shown in a two-dimensional band energy-velocity space.  The grey shading is a guide to the eye.  (e) \& (f) Equal-time band energy-velocity distribution functions for the metallic ($U=0.5t^{*}$) and insulating ($U=2t^{*}$) regimes of Fig. 1(b) and (c), respectively, for various times (units of $1/t^{*}$).  The band energy-velocity space spans a radius of $3.9t^{*}$ with the center of each plot at the origin (
$0t^{*}$, $0t^{*}$) as indicated by the band energy-velocity axes in the top panels.  A time-lapse sequence of images for both cases with finer resolution can be found in the Supplementary Material.\label{fig:4}}
\end{figure}

Consider the results shown in Fig.~\ref{fig:4}.  Plotted versus the band energy $\varepsilon_\mathbf{k}$ and the normalized electron velocity $v_\mathbf{k}=\mathbf{v}_\mathbf{k} \cdot \hat{E}$, the line dividing the occupied and empty states rotates at a rate given by $-\partial \mathbf{A}(t)/\partial t \cdot \hat{E}=\mathbf{E}(t) \cdot \hat{E}$, simply the magnitude of the electric field as a function of time.  One may remove this rotation by going-over to an instantaneous frame which corresponds to examining the distribution functions in a particular gauge where one observes a static Fermi-Dirac distribution with no temporal dynamics.  While the underlying physics remains unchanged, visualizing the results depends on whether one works with the gauged or gauge-invariant functions.  We choose the latter (gauge-invariant formulation) as illustrated for a simple half-filled one-dimensional noninteracting band metal in Figs.~\ref{fig:4}(a-d) and used to display the instantaneous electronic distribution 
functions for the metallic and insulating regimes shown in Figs.~\ref{fig:4}(e) and (f), respectively.

Incorporating the influence of interactions has a number of effects on the temporal behavior.  First, the initial distribution broadens due to the electron-electron interactions, but remains independent of $v_\mathbf{k}$ and monotonically decreases as a function of $\varepsilon_\mathbf{k}$.  The applied field drives particles via the Peierls' substitution, but now electron-electron scattering randomizes the distribution and eventually quenches Bloch oscillations initially observable in the distribution function.  After the pump pulse decays, electron-electron scattering drives the distribution toward a more-or-less static, steady-state, pattern.  If the distribution depends on $v_\mathbf{k}$ or, as primarily observed in these simulations, is no longer monotonically decreasing as a function of $\varepsilon_\mathbf{k}$, then the quasithermal paradigm cannot hold.

Figs.~\ref{fig:4}(e) and (f) show the equal-time distribution function for a sequence of times in the metallic and insulating regimes, respectively.  In Fig.~\ref{fig:4}(e) weak correlations allow the electric field pulse to easily shift the distribution function, similar to the expected behavior for a simple band metal.  The system maintains a well-defined ``Fermi edge'', as observed in equilibrium, for all but those times with the strongest electric field near the center of the pulse ($0/t^{*}$).  The edge reforms as the pump pulse decays and at the longest simulation times a significantly wider edge appears indicative of the higher effective temperatures, and validity of the quasithermal paradigm, used to describe the observed photoemission response in Fig.~\ref{fig:3}.   

In Fig.~\ref{fig:4}(f), electronic correlations on the insulating side of the Mott MIT produce an equilibrium distribution also with a well-defined ``Fermi edge''.  These correlations provide an initial resistance to the influence of the applied pump pulse at short times and significantly scramble the electron redistribution at long times.  Relaxation through electron-electron scattering induces a partial reformation of the edge at the longest simulation times; however, the system retains a significant nonthermal distribution of weight characterized primarily by a nonmonotonic dependence on $\varepsilon_\mathbf{k}$.  A sequence of snapshots for both cases with finer time resolution and snapshots for the observed behavior with an alternative pump pulse can be found in the Supplementary Material.

These results reveal a dichotomy in the evolution of transiently excited electrons as one tunes across the Mott MIT.  The quasithermal picture, which has served to underpin much of the experimental analysis, remains valid in the metallic regime where relaxation can be characterized by evolution toward an effectively thermal steady-state.  Tuning correlations to the insulating regime, across the MIT, causes a breakdown in this paradigm as one clearly observes relaxation toward a nonthermal state.  On short time scales where the electronic system behaves like an isolated system not in contact with additional degrees of freedom, one must take this dichotomy into account when performing experimental analysis.  Additional interactions, not considered here, should dominate the much longer time recovery where the system must naturally return to its original equilibrium through coupling to the crystal lattice (electron-phonon coupling) and eventually ballistic and diffusive transport of the delivered pump energy to 
the material's bulk and subsequently the environment.

\begin{acknowledgments}
The authors would like to thank A. Cavalleri, M. Eckstein, P. S. Kirchmann, M. Kollar, H. R. Krishnamurthy, A. Lindenberg, D. Reis, F. Schmitt, and M. Wolf for valuable discussions.  BM, AFK, MS, and TPD were supported by the U.S. DOE, BES, MSED under Contract No.~DE-AC02-76SF00515.  JKF was supported by the U.S. DOE, BES, MSED under Grant No.~DE-FG02-08ER46542 and by the McDevitt bequest at Georgetown University.  The collaboration was supported by the U.S. DOE, BES through the CMCSN program under Grant No.~DE-SC0007091. This work benefited from an INCITE grant administered by the U.S. DOE, ASCR utilizing the resources of NERSC supported by the U.S. DOE, Office of Science, under Contract No.~DE-AC02-05CH11231.
\end{acknowledgments}

\bibliography{Main_Moritz_Bib}

\pagebreak[4]

\appendix

\renewcommand{\thefigure}{S\arabic{figure}}
\renewcommand{\theequation}{S\arabic{equation}}
\setcounter{figure}{0}
\setcounter{equation}{0}

\begin{center}
\textbf{Supplementary Material:  Correlation tuned cross-over between thermal and nonthermal states following ultrafast transient pumping}
\end{center}

\section{S1:  Scaling}

Scaling of physical quantities in the nonequilibrium DMFT formalism begins by considering the proper energy scale via the noninteracting kinetic energy.  Consider only nearest-neighbor hopping on the infinite dimensional hypercubic lattice.  Then the band energy $\varepsilon_{\mathbf{k}}=\lim_{d\rightarrow\infty} -2t\sum_{i=1}^{d}\cos(k_{i})$, where $d$ is the lattice dimension.  The hopping is chosen to scale$^{\textrm{S}1}$ as $t=t^{*}/2\sqrt{d}$ and we take $t^{*}$ as the base energy unit for the problem.  This scaling leads to a Gaussian noninteracting density of states and a finite effective bandwidth.  Since the direction of the applied field was chosen along the hypercubic body diagonal, the Pierels' substitution natural produces a second ``band energy'' given by $\overline{\varepsilon}_{\mathbf{k}} = \lim_{d\rightarrow\infty} -\left(t^{*}/\sqrt{d}\right)\sum_{i=1}^{d}\sin(k_{i})$, which disappears from the problem in the limit where the strength of the applied field goes to zero.  These two band 
energies have a Gaussian joint density of states which appears in the Hilbert transform used to obtain the local Green's function in the iterative DMFT process.

With the effective energy scale $t^{*}$, time is measured in units of $1/t^{*}$ and the magnitude of the effective field would have units of $t^{*}/e$.  To convert these to some physical scale, first we choose $t^{*}\sim0.1$ eV.  Our fundamental unit of time becomes $\hbar/t^{*}\sim6.6$ fs; and if we assume the characteristic length scale $a_{\textrm{o}}\sim3$\AA{}, the characteristic unit for the field $E_{\textrm{o}}=t^{*}/ea_{\textrm{o}}\sim33$ mV/\AA{}.  To add some additional perspective, the equilibrium temperature for all cases considered in the main text would be $T=0.1t^{*}\sim116$ K; and for the pump pulse in the main text $\omega_{\textrm{p}}=0.5t^{*}\sim76$ THz, $\sigma_{\textrm{p}}=5/t^{*}\sim33$ fs, and $E_{\textrm{max}}\sim800$ mV/\AA{} (a large value, but not unrealistic given the field strengths currently employed in some pump-probe experiments).

\section{S2:  Power and Energy}

\begin{figure}[!t]
\includegraphics[width=\columnwidth]{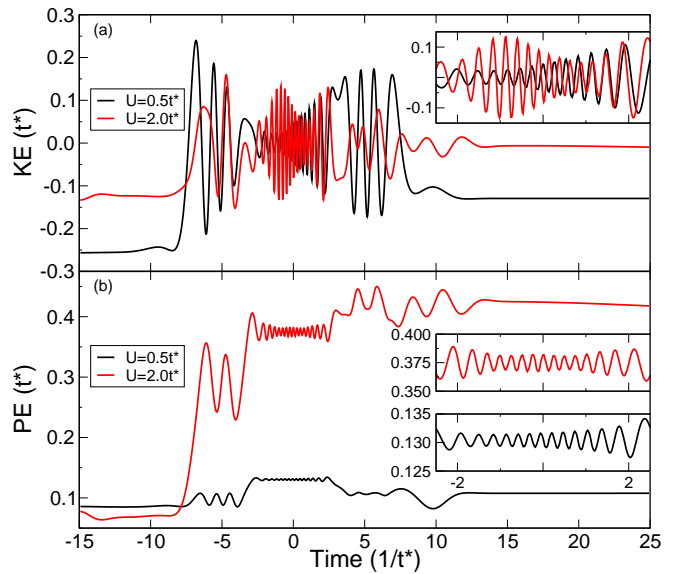}
\caption{(a) The kinetic energy (KE) as a function of time for the two interaction regimes in the main text. (b) The corresponding potential energy (PE) as a function of time.  The insets highlight those times near the center of the pump pulse where one sees rapid variation in the KE as a function of time.  The effect of the pump pulse manifests as an increase in KE in both the metallic and insulating regimes; however, the PE shows a distinct difference, as one naively expects, with a significant PE increase as a function of time in the insulating regime due to the applied pump pulse.
\label{fig:s1}}
\end{figure}

We evaluate the instantaneous power, kinetic energy (KE), potential energy (PE), and total energy (see Fig.~2 in the main text and Fig.~\ref{fig:s1}).  The instantaneous power delivered by the pump pulse can be determined in a straightforward manner by first evaluating the instantaneous current 
\begin{equation}
\left<\mathbf{j}(t)\right>=-ei\sum_{\mathbf{k}}\mathbf{v}_{\mathbf{k}-e\mathbf{A}(t)} G^{<}_{\mathbf{k}}(t,t), 
\end{equation}
or alternatively, as we do in this case, one can express both the band velocity and momentum-dependent lesser Green's function in terms of the two band energies and convert the sum over momentum into a weighted integral over $\{\varepsilon,\overline{\varepsilon}\}$ as one does in evaluating the local Green's function.$^{\textrm{S}2}$  Then the instantaneous power $P(t) = \left<j(t)\right>E(t)$, as both the current and field point along the hypercubic body-diagonal in our simulations.  Since we are dealing with a closed system, the integral of the instantaneous power as a function of time gives the total energy delivered by the applied field.  

We evaluate the KE in much the same fashion as the instantaneous current,$^{\textrm{S}3}$ since one can merely replace the velocity $\mathbf{v}_{\mathbf{k}}$ with the band energy $\varepsilon_{\mathbf{k}}$ and drop the electric charge (again one equivalently performs an integral over $\{\varepsilon,\overline{\varepsilon}\}$ in the numerical evaluation).  The total 
system energy is determined 
from the time derivative of the local lesser Green's function$^{\textrm{S}3}$
\begin{equation}
\textrm{Total Energy} = \left.\frac{\partial G^{<}_{\textrm{loc}}(t,t')}{\partial t}\right|_{t'=t},
\end{equation}
analogous to the equilibrium expression.  While we could, in principle, evaluate the PE from a product of the self-energy and Green's function analogous to the equilibrium expression,$^{\textrm{S}3}$ the system is closed and the PE = Total Energy - KE.  Each of these quantities has been extrapolated independently to the limit of zero discretization on the Keldysh contour.

\section{S3:  Quasithermal Fitting}

\begin{figure}[!t]
\includegraphics[width=\columnwidth]{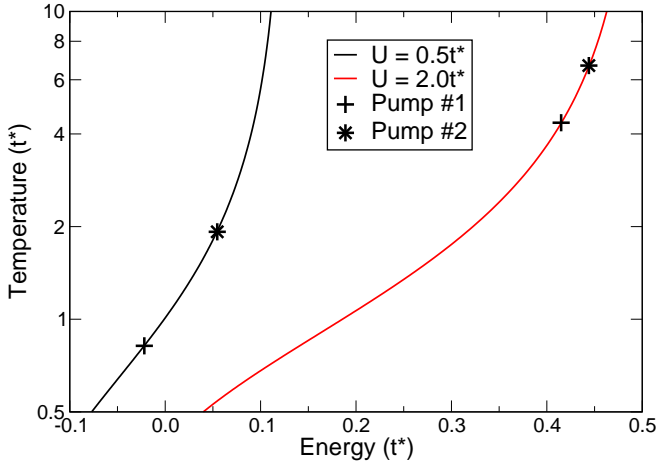}
\caption{The temperature T versus the total system energy in the Falicov-Kimball model for the two values of interaction $U$ studied in the main text.  Points highlight the energy-temperature values corresponding to the pump pulse from the main text (Pump \#1) and a second (Pump \#2) appearing only in this Supplementary Material with characteristics $E_\textrm{max}$=50$E_\textrm{o}$, $\omega_\textrm{p}=1.0t^{*}$, and $\sigma_\textrm{p}=5/t^{*}$ (see Fig.~\ref{fig:s5} for a comparison).
\label{fig:s2}}
\end{figure}

For a given interaction strength, the local density of states $N(\omega)$ in the Falicov-Kimball model is temperature independent.  Therefore, in a quasithermal fit to the photoemission response, the temperature enters only in the effective Fermi-Dirac distribution function which gives the effective local lesser Green's function $-iG^{<}_{\textrm{loc}}(\omega)=N(\omega)f(\omega,T)$.  One measure for quasithermal effective temperature comes from the total energy (see Fig.~2 in the main text).  Figure~\ref{fig:s2} shows the temperature as a function of total energy for the two interaction strengths discussed in the main text.  Here, the total energy for a given temperature has been determined from the first moment of $N(\omega)f(\omega,T)$.  Points have been included in Fig.~\ref{fig:s2} for two different pump pulses: the one listed in the main text (Pump \#1) and a second (Pump \#2) discussed only in the Supplementary Material.  Using these effective temperatures, the quasithermal photoemission response can 
be determined straightforwardly.$^{\textrm{S}4}$  The results appear as solid lines in Fig.~3 of the main text or Fig.~\ref{fig:s4} for the two pump pulses, respectively.

\begin{figure}[!t]
\includegraphics[width=\columnwidth]{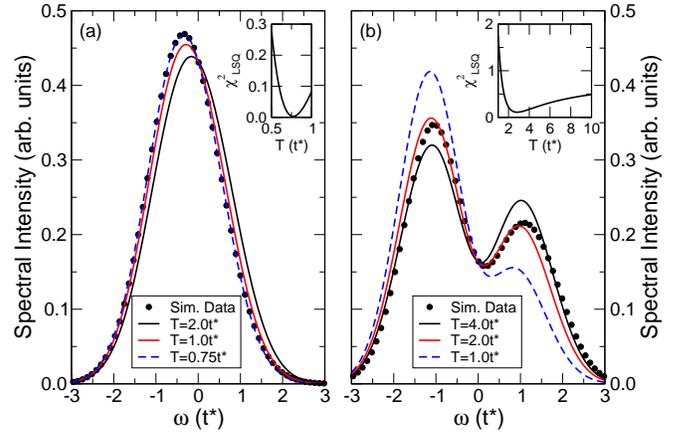}
\caption{The equilibrium photoemission response for [a, $U=0.5t*$] metallic and [b, $U=2.0t*$] insulating regimes discussed in the main text for several different temperatures T.  The insets show the least-squares fit to the simulated response function at long times used to determined the least-squares ``best-fit'' response appearing in Fig.~3(a) and (b), respectively.  Black dots represent data from the simulation as in the main text (see Fig.~3), shown only for every tenth data point for clarity.  The full data set has been used to determine the least squares fit in each case.  
\label{fig:s3}}
\end{figure}

We also have determined the effective quasithermal response for a large range of temperatures.  Figure~\ref{fig:s3} shows some representative results for a few temperatures compared to the simulated data.  The inset in each panel of Fig.~\ref{fig:s3} shows the sum of the square difference between the quasithermal response and the simulation data as a function of temperature.  The quasithermal response for the temperature corresponding to the minimum $\chi_{\textrm{LSQ}}^{2}$ in each case determines the least-squares fit appearing in Fig.~3(a) and (b) of the main text, respectively.  A similar analysis has been conducted for the second pump pulse (P. \#2) appearing in Fig.~\ref{fig:s4}. 

\begin{figure}[!t]
\includegraphics[width=\columnwidth]{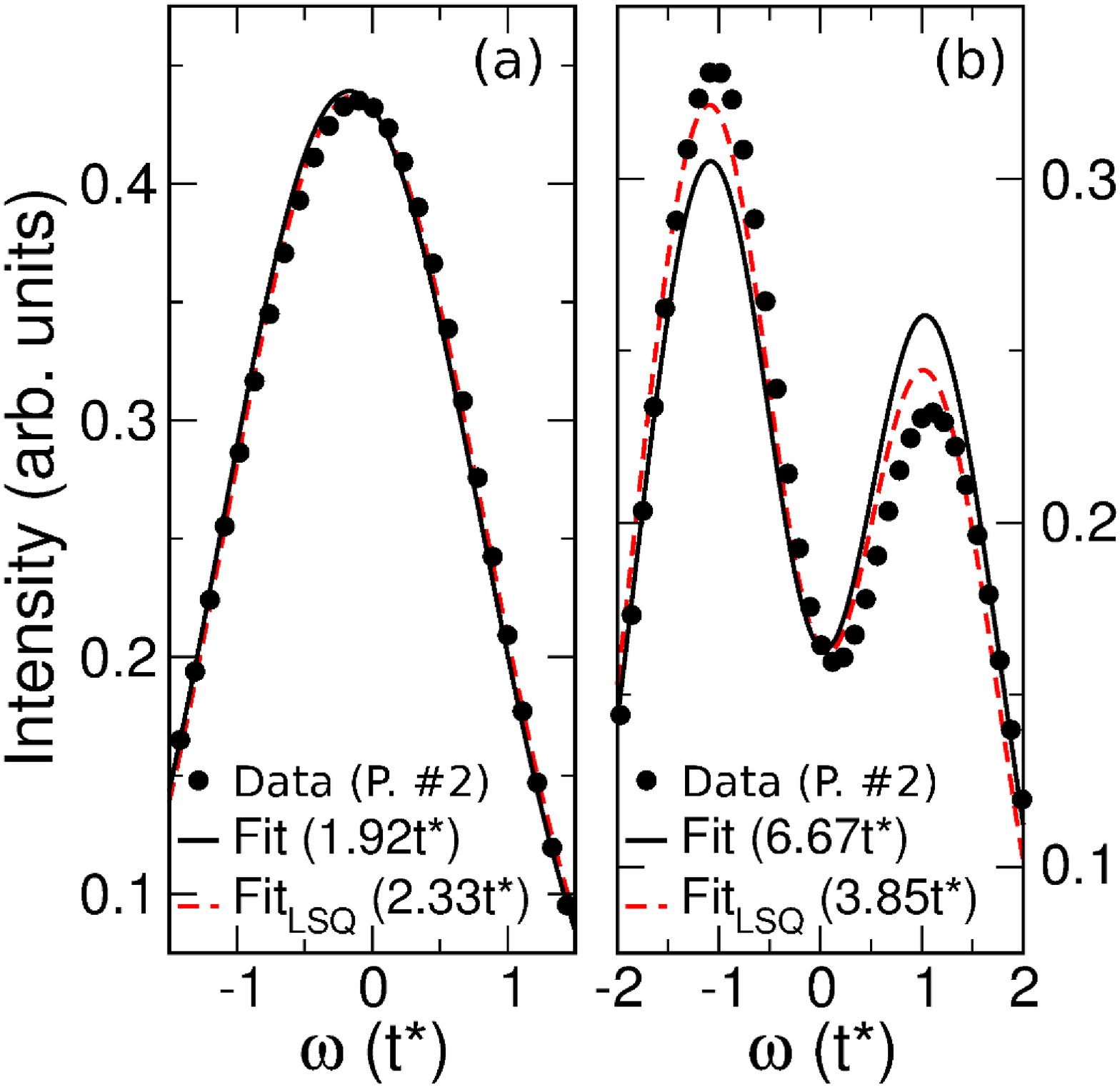}
\caption{Quasithermal fits of the steady-state photoemission response (characteristic probe width $\sigma_\textrm{b}=2/t^{*}$) for (a) metallic ($U=0.5t^{*}$) and (b) insulating ($U=2t^{*}$) systems subject to Pump \#2 (P. \#2), respectively, with the effective temperature shown in parenthesis.  The fits (solid black) are determined by extracting the temperature from the total energy, while the best-fit quasithermal response (dashed red) has determined by the least-squares method as described in Section S3.  
\label{fig:s4}}
\end{figure} 

\begin{figure}[!th]
\includegraphics[width=\columnwidth]{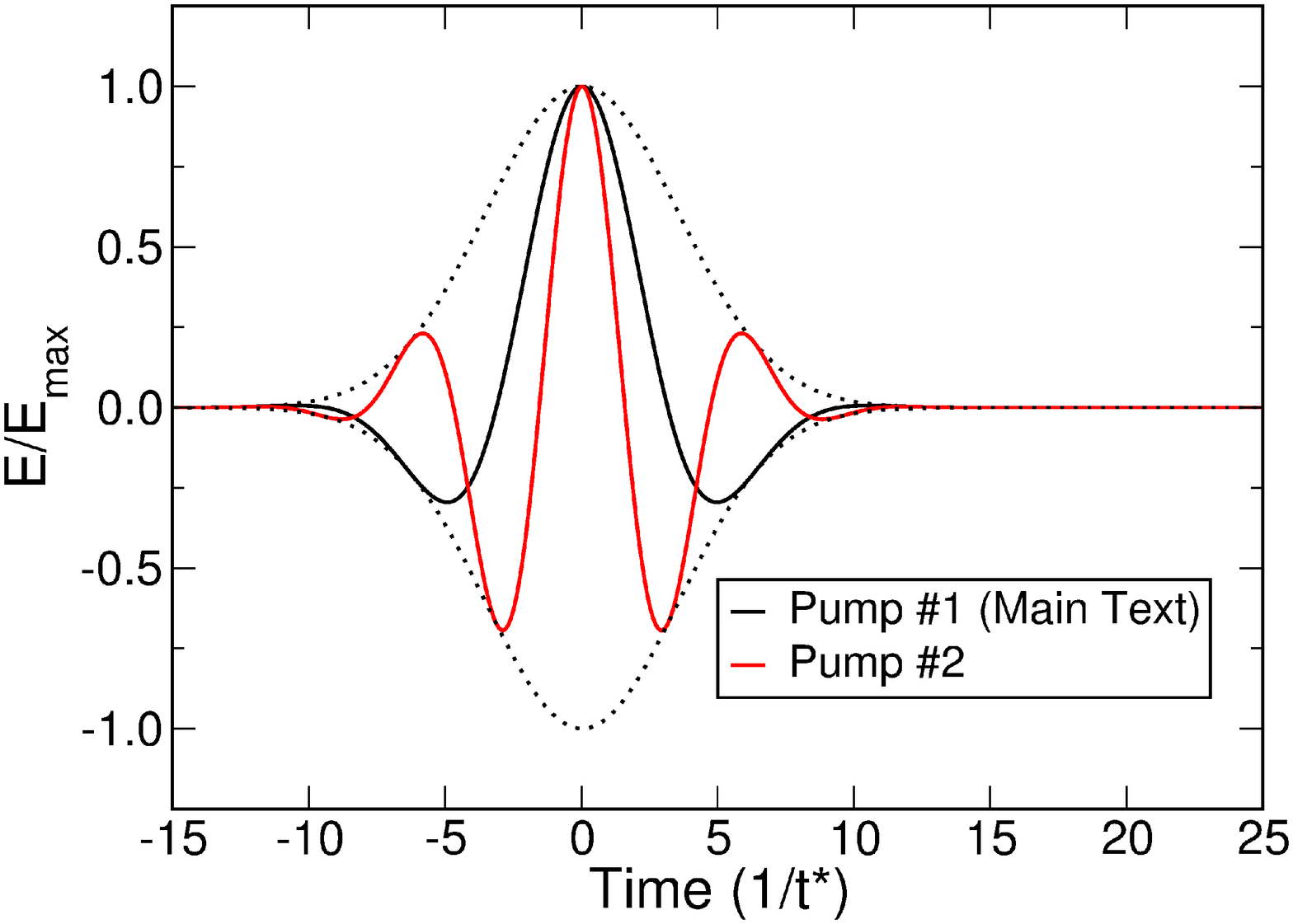}
\caption{Temporal profile of the electric field pump pulse $\mathbf{E}$ for two different cases:  the one discussed in the main text (Pump \#1) and the second at higher fundamental frequency and field strength (Pump \#2) as described in Fig.~\ref{fig:s2}.  
\label{fig:s5}}
\end{figure} 

\section{S4:  Movies}

As stated in the caption to Fig.~4 in the main text a series of movies with fine-grained time resolution have been produced for both the (a) metallic and (b) insulating regimes under study subject to the two pump pulses.  Figure~\ref{fig:s5} shows the temporal profile of the electric field $\mathbf{E}$ for the two cases.  The notation used for the movies is SM followed by a numeral representing the pump number and \emph{m} or \emph{i} representing the metallic or insulating regime, respectively, \emph{e.g.} the movie following the time evolution of the equal-time distribution function for the metallic system subject to the pump discussed in the main text (Pump \#1) is SM1m and that following the evolution for the insulating system subject to the second pump (Pump \#2) is SM2i.    As in the main text, the band energy-velocity space spans a radius of $3.9t^{*}$ with the center at the origin ($0t^{*}$, $0t^{*}$) as indicated by the axis at the center of the plots.  The respective time for the frames of each 
movie has been annotated in the upper left hand corner in units of $1/t^{*}$ with the interaction strength $U$ in units of $t^{*}$ on the right.  In each case, the distribution function ``rotates'' in band energy-velocity space.  In the metallic regime, the ``Fermi edge'' present in equilibrium reforms at long times with a significantly wider distribution indicative of the increase in effective temperature; at long times one observes the nonmonotonic dependence on $\varepsilon_{\mathbf{k}}$ in the insulating regime indicative of a nonthermal state following the ultrafast transient pumping.  

\section{Supplemental References}

\noindent S1. W. Metzner and D. Vollhardt, {\it Phys.\ Rev.\ Lett.} \textbf{62}, 324-327 (1989).

\noindent S2. J. K. Freericks, {\it Phys.\ Rev.\ B} \textbf{77}, 075109 (2008).

\noindent S3. V. M. Turkowski and J. K. Freericks, {\it Phys.\ Rev.\ B} \textbf{73}, 075108 (2006).

\noindent S4. J. K. Freericks, H. R. Krishnamurthy, and T. Pruschke, {\it Phys.\ Rev.\ Lett.} \textbf{102}, 136401 (2009).

\vfill

\end{document}